\documentclass[aip,nofootinbib]{revtex4-1}
\usepackage{amssymb}
\usepackage{rotating}
\usepackage{xypic}
\usepackage{amssymb}
\usepackage{amsmath,amscd}
\usepackage[english]{babel}
\usepackage{pst-node,pstricks,multido,pst-plot,pst-text,pst-3d}
\usepackage{graphicx}
\usepackage{amsfonts}
\usepackage[latin1]{inputenc}
\newtheorem{example}{Example}[section]

\newtheorem{remark}[example]{Remark}
\newtheorem{theorem}[example]{Theorem}

\newtheorem{corollary}[example]{Corollary}
\newtheorem{conjecture}[example]{Conjecture}

\newtheorem{proposition}[example]{Proposition}

\newtheorem{lemma}[example]{Lemma}

\def\cal#1{{\mathfrak #1}}

\def\<{\langle}
\def\>{\rangle}

\def\C{{\mathbb C}}

\def\N{{\mathbb N}}

\def\ie{{\emph i.e.}, }

\def\ashuff#1#2#3{
\kern 1pt \vrule height#1 \overline{\vrule height#3 width 0pt
\hskip#2} \rule{.3pt}{#1}\overline{\vrule height#3 width 0pt
\hskip#2} \rule{.3pt}{#1} \kern 1pt }

\def\wt#1{#1}
\def\Tr{{\rm Tr}}

\begin{document}
\title[Selberg-like integrals]{Asymptotics of  Selberg-like integrals:\\ The unitary case and  Newton's interpolation formula}
\author{Christophe Carr\'e}\affiliation{LITIS - D\'epartement d'informatique de
l'Universit\'e de Rouen.\\ Avenue de l'Universit\'e - BP 8 76801
Saint Etienne du Rouvray, France}
\author{Matthieu Deneufchatel}
\affiliation{Laboratoire d'Informatique de Paris-Nord
UMR CNRS 7030
Institut Galil\'ee - Universit\'e Paris-Nord
99, avenue Jean-Baptiste Cl\'ement 
93430 Villetaneuse, France}
\author{Jean-Gabriel Luque}
\affiliation{LITIS - D\'epartement d'informatique de
l'Universit\'e de Rouen.\\ Avenue de l'Universit\'e - BP 8 76801
Saint Etienne du Rouvray, France}
\author{Pierpaolo Vivo}
\affiliation{ICTP - Abdus Salam International Centre for Theoretical Physics\\
Strada Costiera 11, 34151 Trieste, Italy}
\begin{abstract}
We investigate the asymptotic behavior of the Selberg-like integral
\[
 \frac1{N!}\int_{[0,1]^N}x_1^p\prod_{i<j}(x_i-x_j)^2\prod_ix_i^{a-1}(1-x_i)^{b-1}dx_i,
\]
as $N\to\infty$ for different scalings of the parameters $a$ and $b$ with $N$. Integrals of this type arise in the random matrix theory of electronic scattering in chaotic cavities supporting $N$ channels in the two attached leads.
Making use of Newton's interpolation formula, we show that an asymptotic limit exists and we compute it explicitly.
\end{abstract}

 \maketitle

\section{Introduction}
In his famous 1944 paper \cite{Selberg}, Atle Selberg introduced
and computed the integral
\begin{equation}\begin{array}{r}\displaystyle
 S_{N}(a,b,\beta):=\int_{[0,1]^N}\prod_{1\leq i<j\leq N}|x_i-x_j|^{2\beta}\prod_{i=1}^Nx_i^{a-1}(1-x_i)^{b-1}dx_i
 \\\displaystyle=\prod_{j=0}^{N-1}{\Gamma(a+j\beta)\Gamma(b+j\beta)\Gamma(1+(j+1)\beta)\over \Gamma(a+b+(N+j-1)\beta)
 \Gamma(1+\beta)}\end{array}
\end{equation}
in the aim to solve a problem of Gelfond \cite{Gelfond}. Since the
sixties, many generalizations and applications have been developed
(see \cite{FW} for an interesting review about  Selberg
integral). The scope of these investigations involves many areas
of mathematics :  random matrices \cite{Meh,For},  calculations
of constant terms (see {\it e.g.} \cite{Dy,Good}), symmetric
functions \cite{lasc,Macd} (in particular, Jack and Macdonald
polynomials \cite{Kan,Kor}), multivariate orthogonal polynomials
\cite{Las}, the value distribution of the Riemann $\zeta$
function on the critical line \cite{KN} among other applications. Since the
seventies, many applications in physics have been found,
especially in the theory of quantum Hall effect
\cite{Lau,DGIL}. The number of variables involved in the integrand
is then interpreted as the number of particles. In this context it
is interesting to study what happens when this number becomes very large.

In the field of random matrices, the integrand of Selberg's integral
corresponds also (for quantized values of $\beta$) to the joint
probability density of eigenvalues of one of the classical random
matrix ensembles, the Jacobi ensemble. Matrices from this ensemble
can be generated in three different ways:
\begin{itemize}
\item as truncations of Haar orthogonal, unitary or symplectic
matrices \cite{sommerstrunc}. For the case of unitary matrices, an
important application arises in the theory of electronic transport
in mesoscopic systems at low temperatures \cite{beenakker}, as
detailed below. \item as composition of Wishart matrices, with
applications to multivariate statistics \cite{muir}. \item as
composition of projection matrices \cite{collins}.
\end{itemize}
In the theory of quantum transport through mesoscopic devices
(Landauer-B\"{u}ttiker scattering approach
\cite{beenakker,landauer,buttikerPRL}), the wave function
coefficients of the incoming and outgoing electrons in a cavity
are related through the unitary scattering matrix $S$ ($2N\times
2N$, if $N$ is the number of electronic channels that each lead
supports):
\begin{equation}\label{ScatteringMatrix S}
  S=\left(
  \begin{array}{cc}
    r & t^\prime \\
    t & r^\prime
  \end{array}\right)
\end{equation}
where the transmission ($t,t^\prime$) and reflection
$(r,r^\prime)$ blocks are $(N\times N)$ matrices encoding the
transmission and reflection coefficients among different channels.
Many quantities of interest for the experiments are \emph{linear
statistics} on the eigenvalues of the hermitian matrix $t
t^\dagger$, \ie are quantities of the form $A=\sum_{i=1}^N
f(T_i)$ where $f(x)$ is a smooth function (not necessarily linear)
and $T_i$ are the eigenvalues of $t t^\dagger$ (real numbers
 between $0$ and $1$), which have the intuitive interpretation as the probability
that an electron gets transmitted through the $i$-th channel. 
 For
example, the dimensionless conductance and the shot noise are
given respectively by $G=\Tr(t t^\dagger)$ \cite{landauer} and
$P=\Tr[t t^\dagger(1-t t^\dagger)]$ \cite{lesovik,ya}. The random
scattering theory models the scattering matrix $S$ for the case of
chaotic dynamics as a random unitary matrix uniformly distributed
in the unitary group, \ie it belongs to one of Dyson's Circular
Ensembles.

From this information, the joint probability density of the
transmission eigenvalues $\{T_i\}$ of the matrix $t t^\dagger $,
from which the statistics of interesting experimental quantities
could be in principle derived, is readily recognized as the
Selberg integrand (Jacobi measure) with $b=1$ and $2\beta=1,2,4$
\cite{forrcond} depending on physical symmetries of the Hamiltonian (for recent results on the use of 	Selberg integral in the quantum transport problem, see \cite{sommers,savinnew,savin,savin2}).
 The importance of linear statistics, and their
asymptotical properties when the number of channels grows to
infinity, provides one of our main motivations for the present
study.

 This paper is the continuation of \cite{LV} and we are interested in the following integrals
\[
 \langle
 f(x_1,\dots,x_N)\rangle_{a,b}=\frac1{N!}\int_{[0,1]^N}f(x_1,\dots,x_N)\prod_{i<j}(x_i-x_j)^2\prod_ix_i^{a-1}(1-x_i)^{b-1}dx_i,
\]denoting  averages over the Jacobi probability density.
Especially when $f$ is a power sum (section \ref{sectionpowersum})
$$
 p_k:=\sum_i x_i^k,
$$
 or a Schur function (section \ref{sectionschur}), we raise the question of its asymptotic behavior
for $N\to\infty$.

More precisely, using classical identities on symmetric functions
we show  that this problem reduces to the calculation of an
inverse binomial transform.\\
This paper is the first step towards a combinatorial
interpretation of the asymptotic behavior of Selberg-like
integrals. Note that J.-Y. Thibon with one of the authors have
already investigated the links between combinatorics and Selberg
integrals \cite{LT1,LT2,LT3}.

The plan of the paper is as follows. In section \ref{Integ}, we
give an expression of the integrals as a rational function in the
numbers of variables $N$. In section \ref{binom}, we investigate
some properties of the binomial transform which will be used to
compute the limit values in section \ref{Asymp}. Section
\ref{Combi} deals with  several special cases related to 
combinatorics. Finally, in section \ref{conclusion} we provide concluding remarks,
and in appendix \ref{spectral} an alternative approach to corollary \ref{I_k}.

\section{Some Selberg-like integrals \label{Integ}}
\subsection{Schur functions\label{sectionschur}}
Here we give an expression of
 $\langle
s_\lambda(x_1,\dots,x_N)\rangle_{a,b} \over\langle
1\rangle_{a,b}$, where
$$s_\lambda(x_1,\dots,x_N)={\det(x_i^{\lambda_i+N-j})_{1\leq
i,j\leq N}\over \prod_{1\leq i<j\leq N}(x_i-x_j)}$$ denotes a
Schur function (see {\it e.g.} \cite{Macd,lasc}). \\
A rather classical formula which can be found in the book of
Macdonald \cite{Macd}(this is a special case of the exercice 7
p385) gives
\[
\langle
s_\lambda(x_1,\dots,x_N)\rangle_{a,b}=\prod_{i<j}(\lambda_i-\lambda_j+j-i)\prod_{i=1}^N{\Gamma(\lambda_i+a+N-i)\Gamma(b+N-i)\over
\Gamma(\lambda_i+a+b+2N-i-1)}.
\]
Hence,
\[\begin{array}{rcl}\displaystyle
{\langle s_\lambda(x_1,\dots,x_N)\rangle_{a,b} \over\langle
1\rangle_{a,b}}&=&\displaystyle\prod_{i<j}{\lambda_i-\lambda_j+j-i\over
j-i}\times\\&&\times\displaystyle
\prod_{i=1}^N{\Gamma(\lambda_i+a+N-i)\over \Gamma(a+N-i)}{
\Gamma(a+b+2N-i-1)\over\Gamma(\lambda_i+a+b+2N-i-1)}.
\end{array}\]

First, we remark that the number of factors of
\[
\prod_{i=1}^N{\Gamma(\lambda_i+a+N-i)\over \Gamma(a+N-i)}{
\Gamma(a+b+2N-i-1)\over\Gamma(\lambda_i+a+b+2N-i-1)} =
 \prod_{i=1}^{\ell(\lambda)}\prod_{j=0}^{\lambda_i-1}{a+N-i+j\over
 a+b+2N-i+j-1},
\]
depends only on the partition $\lambda$ and not on the number of
variables $N$. In the same way, one has
\[
\prod_{i<j}{\lambda_i-\lambda_j+j-i\over
j-i}=\prod_{i=1}^{\ell(\lambda)}\left[\prod_{j=i+1}^{\ell(\lambda)}
{\lambda_i-\lambda_j+j-i\over
j-i}\prod_{j=\ell(\lambda)+1}^N{\lambda_i+j-i\over j-i} \right].
\]
One needs the following lemma which is obtained by reorganizing
the factors and simplifying the resulting expression.
 \begin{lemma}\label{lemmFact1}
  For $N$ large enough\footnote{The condition ``{\it $N$ large enough}'' can be omitted provided that we use the notation $\prod_{i=a}^bf(i)=\prod_{i=b}^af(i)^{-1}$.}, one has
  \[
  \prod_{i=a+1}^N{b+i\over b+c+i}=\prod_{i=1}^{c}{a+b+i\over b+N+i}
  \]
  where $a,c\in\N$.
 \end{lemma}
 {\bf Proof}
 First write
 {\footnotesize
 \begin{equation}\label{fact1}
 \begin{array}{rcl}
\displaystyle\prod_{i=a+1}^N{b+i\over
b+c+i}&=&\displaystyle\prod_{i=0}^{N-a-1}{b+i+a+1\over
b+c+i+a+1}\\
&=&{\displaystyle
\prod_{i=0}^{c-1}(b+i+a+1)\prod_{i=c}^N(b+i+a+1)\over\displaystyle
\prod_{i=0}^{N-a-c-1}(b+c+i+a+1)\prod_{i=N-c-a}^{N}(b+c+i+a+1)}.
\end{array}
\end{equation}
 }
 But
 \[\prod_{i=c}^N(b+i+a+1)=\prod_{i=0}^{N-c}(b+c+i+a+1),\]
 and
\[\prod_{i=N-c-a}^{N}(b+c+i+a+1)=\prod_{i=0}^{c-1}(b+N+i+1).\]
Hence, by substituting these two identities in (\ref{fact1}), one
recovers the result.$\Box$.\\ \\
If one applies lemma \ref{lemmFact1} to
$\prod_{j=\ell(\lambda)+1}^N{\lambda_i+j-i\over j-i}$, one finds
\[
\prod_{j=\ell(\lambda)+1}^N{\lambda_i+j-i\over
j-i}=\prod_{j=1}^{\lambda_i}{j+N-i\over \ell(\lambda)+j-i}.
\]
Hence,
\begin{proposition}\label{Schur}
One has {\footnotesize
\begin{equation}\label{SchurEq}\begin{array}{l}
\displaystyle{\langle s_\lambda(x_1,\dots,x_N)\rangle_{a,b}
\over\langle
1\rangle_{a,b}}=\\\displaystyle\prod_{i=1}^{\ell(\lambda)}\left[\prod_{j=i+1}^{\ell(\lambda)}
{\lambda_i-\lambda_j+j-i\over
j-i}\prod_{j=0}^{\lambda_i-1}{(j+N-i+1)(a+N-i+j)\over
(\ell(\lambda)+j-i+1)( a+b+2N-i+j-1)} \right]
\end{array}\end{equation}} If $a=a_1N+a_0$ and $b=b_1N+b_0$ are two
linear functions of $N$, this implies that
{\footnotesize\[\begin{array}{l}\displaystyle{\langle
s_\lambda(x_1,\dots,x_N)\rangle_{a,b} \over N^{|\lambda|}\langle
1\rangle_{a,b}}\mathop\rightarrow_{N\rightarrow\infty}
\left(\frac{1+a_1}{2+a_1+b_1}\right)^{|\lambda|}\displaystyle\prod_{i=1}^{\ell(\lambda)}\left[\prod_{j=i+1}^{\ell(\lambda)}
{\lambda_i-\lambda_j+j-i\over j-i}\prod_{j=0}^{\lambda_i-1}{1\over
\ell(\lambda)+j-i+1} \right] ,\end{array} \]} where
$|\lambda|=\sum_i\lambda_i.$
\end{proposition}

 \subsection{Selberg-like integrals with a power sum in the integrand\label{sectionpowersum}} In this section, we study the integral
\[
 I_k:={\langle p_k(x_1,\dots,x_N)\rangle_{a,b}\over \langle
 1\rangle_{a,b}},
\]
where $p_k=\sum_ix_i^k$.

One uses the formula (see {\it e.g.} \cite{lasc,Macd}):
\begin{equation}\label{ptos}
p_k=\sum_{i=0}^{k-1}(-1)^is_{[(k-i)1^i]},
\end{equation}
where $[(k-i)1^i]$ denotes the partition $[(k-i),\underbrace{1,\dots,1}_{\times i}]$.
Hence,
\begin{equation}\label{expa1}
I_k=\sum_{i=0}^{k-1}(-1)^i{\langle
s_{[(k-i)1^i]}\rangle_{a,b}\over\langle 1\rangle_{a,b}}.
\end{equation}
From proposition \ref{Schur}, one has
\begin{corollary}\label{I_k}
For each $k>0$, one has
\[
I_k=\frac1{k!}\sum_{i=0}^{k-1}(-1)^i\left(k-1\atop
i\right)\prod_{j=-i}^{k-i-1}{(N+j)(a+N+j-1)\over a+b+2N+j-2}.
\]
\end{corollary}
{\bf Proof} From Eq (\ref{SchurEq}), one has {\footnotesize
\begin{equation}\label{equerre}
\begin{array}{rcl}
\displaystyle{\langle s_{[(k-i)1^i]}\rangle_{a,b}\over\langle
1\rangle_{a,b}}&=&\displaystyle\underbrace{\displaystyle
\prod_{p=2}^{i+1}{k-i+p-2\over
p-1}\prod_{p=0}^{k-i-1}{(p+N)(a+N-1+p)\over
(i+p+1)(a+b+2N+p-2)}}_{\mbox{first part of the
partition}}\times\\&&\times\displaystyle\underbrace{\displaystyle\prod_{j=2}^{i+1}{(N-j+1)(a+N-j)\over(i-j+2)(a+b+2N-j-1)}}_{\mbox{other
parts of the partition}}\end{array} \end{equation} }But,
\[
\prod_{p=2}^{i+1}{k-i+p-2\over p-1}=\prod_{p=0}^{i-1}{k-i+p\over
p+1}=\left(k-1\atop i\right),
\]

\[
\prod_{p=0}^{k-i-1}{1\over (i+p+1)}\prod_{j=2}^{i+1}{1\over
(i-j+2)}=\prod_{p=i+1}^{k}{1\over p}\prod_{j=1}^{i}{1\over
i}={1\over k!},
\]
and
\[
\prod_{j=2}^{i+1}{(N-j+1)(a+N-j)\over
a+b+2N-j-1}=\prod_{j=-i}^{-1}{(N+j)(a+N+j-1)\over a+b+2N+j-2}.
\]
Using these equalities in (\ref{equerre}), one finds
\[
{\langle s_{[(k-i)1^i]}\rangle_{a,b}\over\langle 1\rangle_{a,b}}=
\frac1{k!}\left(k-1\atop
i\right)\prod_{j=-i}^{k-i-1}{(N+j)(a+N+j-1)\over a+b+2N+j-2}. \]
 The result is obtained by replacing  $\langle s_{[(k-i)1^i]}\rangle_{a,b}\over\langle 1\rangle_{a,b}$ with its value in
(\ref{expa1}).$\Box$\\ \\
The expression in corollary \ref{I_k} is the starting point in computing the limit
$\lim_{N\to\infty}{I_k\over N}$ (in section \ref{Asymp}), using the tools we are going to 
introduce in the following section.

\section{Inverse binomial transform\label{binom}}
\subsection{Inverse binomial transform and  Newton's interpolation formula}
In this section, we shall use widely the {\it inverse  binomial
transform} (see {\it e.g. }\cite{Barry,Weisstein1}) that operates
on a sequence of polynomials ${\mathbb F}=(f_i(x))_{i}$ by
\[
{\cal B}_k^{-1}[{\mathbb F}]=\sum_{i=0}^k(-1)^{k-i}\left(k\atop
i\right)f_i(x)
\]

\begin{proposition}\label{Stirling}

Let $F(y)=\sum_{i=0}^p\alpha_i(x)y^i$  be the unique
polynomial in $y$ with coefficients in $\C[x]$ of degree $p$
(in $y$) interpolating  the points
$$ (0,f_0(x))\ldots (p,f_p(x)).$$
We have:
\[
{\cal B}_k^{-1}[{\mathbb F}]=k!\sum_{i=k}^p S_{i,k} \alpha_i(x),
\]
where $S_{i,k}$ is a Stirling number of the second kind.
\end{proposition}
{\bf Proof} By linearity, it suffices to show the result for $F(y)=y^p$.
 We remark that in this case
 \begin{equation}\label{Stirling2}
 {\cal B}_k^{-1}[{\mathbb F}]=\sum_{i=0}^k(-1)^{k-i}\left(k\atop i\right)i^p=k!
 S_{p,k},\end{equation}
 by means of the well known formula
 \[
 S_{p,k}={1\over k!}\sum_{i=0}^k(-1)^{k-i}\left(k\atop
 i\right)i^p.
 \]
  $\Box$

Consider the divided difference operator
$\partial_{y_1y_2}$ acting on the right of any expression $f$
in $\{y_1,y_2\}$ by 
$$
f\partial_{y_1y_2}=  {f^{\sigma_{y_1y_2}}-f\over y_2-y_1}
$$
where ${\sigma_{y_1y_2}}$ permutes $y_1$ and $y_2$ in $f$.
\begin{remark}\label{Matthieu}
   Note that, assuming that $k \leq p$, the degree of $y_{0}^{p} \partial_{y_{0}y_{1}} \dots \partial_{y_{k-1}y_{k}}$ is equal to $p-k$. Therefore, if $g(x,y)$ is a polynomial of degree $p$ in $x$ and $y$, the degree of $\displaystyle{\cal{B}_{k}^{-1}} \left[ (g(x,i))_{i} \right]$ equals $p-k$.
\end{remark}

This operator is the main tool to describe the
Newton interpolation.
Indeed, consider a one variable function $f(y)$ and a set of interpolating
variables $\{y_0, \dots, y_k\}$.
One has 
$$\displaylines{f(y)=
f(y_0)
+ f(y_0) \partial_{y_0y_1}(y-y_0)
+\cdots\hfill\cr\hfill\cdots
+ f(y_0) \partial_{y_0y_1}
\cdots
\partial_{y_{k-1}y_k}
(y-y_0)\cdots (y-y_{k-1})+R(y)\cr}
$$
with $R(y_i)=0$ for each $i=0,\dots,k$.

We will denote
$$
f\partial_{m\dots n}=
f(y_m)\partial_{y_my_{m+1}}
\dots, \partial_{y_ny_{n+1}}|_{y_i=i},
$$
 for each pair of integers $m\leq n$, with the special case
$ f\partial _{m\dots m}=f(m)$.
Set, also,
$\partial _i:=\partial_{i\dots i+1}$.

With this notation, the polynomial $f$ of degree
$n-m$ interpolating  the points 
$$(m,f(m)),\dots, (n,f(n))$$
becomes
$$
f(y)=\sum_{j=m}^n f\partial_{m\dots j}
(y-m)\dots (y-(j-1)).
$$

Remark that the Stirling numbers appear when we write $y^p$ in
 terms of falling factorials, $(y)_k:=y(y-1)\dots(y-k+1)$,
\[
 y^p=\sum_{k=0}^pS_{p,k}(y)_k.
\]
This means that the Stirling numbers are the coefficients  in the Newton interpolation of $y^p$ at $y=0, \dots, p$.
With our notations this reads
\begin{equation}
S_{p,k}=y^p\partial_{0\dots k+1}.
\end{equation}
Hence, by linearity, one obtains immediately the following result from proposition \ref{Stirling} :
\begin{corollary}\label{DiffStirl}

Let $k<p$ be two  integers.
Let $F$  be the unique
polynomial in $y$ with coefficients in $\C[x]$ of degree $p$
(in $y$) interpolating  the points
$$ (0,f_0(x))\ldots (p,f_p(x)).$$
We have:
$$
{\cal B}_k^{-1}[{\mathbb F}]=k!F\partial_{0\ldots p}. \Box
$$
\end{corollary}
When acting by $\partial_{0\dots k+1}$ on $y^p$, 
one observes the following (shifted) induction.

\begin{proposition}\label{DiffShift}
$$
y^p\partial_{0\cdots k+1}=
(y+1)^{p-1}\partial_{0\cdots k}$$
\end{proposition}
{\bf Proof}
Since  
$$
y^p_0\partial_{y_0y_1}\dots\partial_{y_ky_{k+1}}=\sum_{i=0}^{p-1}y_1^{p-i-1}\partial_{y_1y_2}\dots
\partial_{y_ky_{k+1}}y_0^i,
$$
the specialization gives
$$
 y^p\partial_{0\cdots k+1}=
y^{p-1}\partial_{1\cdots k+1}.
$$
We conclude the proof by noting that the coefficients in the Newton interpolation of any $f(y)$ at $y=1,\dots,k+1$ are, respectively, equal to the coefficients in the Newton interpolation of $f(y+1)$ at $y=0,\dots,k$.$\Box$

\subsection{An example of generalized (inverse) binomial transform}
As an application, consider the polynomials
$$P_i^k(x;a,b):=\prod_{j=0}^{k-i-1}(x+j+a)\prod_{j=0}^{i-1}(x-j+b).$$
For simplicity, we will denote $P_i^k:=P_i^k(x;a,b)$ when there is no ambiguity.
\begin{proposition}\label{LeadCoeff}
When $p\leq k$, one has
$$
{\cal B}_{k-p}^{-1}[P_p^k,\ldots, P_k^k]=
\prod_{i=0}^{p-1}(x+b-i)\prod_{i=0}^{k-p-1}(b-a-p-i)$$
\end{proposition}
{\bf Proof}
First remark that $y_0^i\partial_{y_0y_1}\dots\partial_{y_jy_{j+1}}$ is a symmetric polynomial in $\{y_0,\dots, y_{j+1}\}$. Hence, we can permute the variables in the expression and obtain
\[
 y_0^i\partial_{y_0y_1}\dots\partial_{y_jy_{j+1}}=y_j^i\partial_{y_jy_{j-1}}\dots\partial_{y_1y_0}\partial_{y_0y_{j+1}}.
\]
Applying the same argument to $y_j^i\partial_{y_jy_{j-1}}\dots\partial_{y_1y_0}$ which is symmetric in the variables  $\{y_0,\dots, y_{j}\}$ one gets
\[
y_j^i\partial_{y_jy_{j-1}}\dots\partial_{y_1y_0}\partial_{y_0y_{j+1}}=y_0^i\partial_{y_0y_{1}}\dots\partial_{y_{j-1}y_j}\partial_{y_0y_{j+1}}.
\]
By definition of $\partial_{y_0y_{j+1}}$ one obtains
\begin{equation}\label{complete1}
 y_0^i\partial_{y_0y_1}\dots\partial_{y_jy_{j+1}}={y_{j+1}^i\partial_{y_{j+1}y_1}\partial_{y_1y_2}\dots \partial{y_{j-1}y_j}-y_0^i\partial_{y_0y_{1}}\dots\partial_{y_{j-1}y_j}\over y_{j+1}-y_0}.
\end{equation}
Again,
\[
 y_{j+1}^i\partial_{y_{j+1}y_1}\partial_{y_1y_2}\dots \partial{y_{j-1}y_j}=y_1^i\partial_{y_1y_2}\dots\partial_{y_jy_{j+1}},
\]
and eq. (\ref{complete1}) becomes
\begin{equation}\label{complete2}
y_0^i\partial_{y_0y_1}\dots \partial_{y_jy_{j+1}}={y_1^i\partial_{y_1y_2}\dots\partial_{y_jy_{j+1}}-y_0^i\partial_{y_0y_{1}}\dots\partial_{y_{j-1}y_j}\over y_{j+1}-y_0}.
\end{equation}
Let us prove the result by induction on $k$. Note that if $k=p$, the result is straightforward. Denote by ${\bf P}(y)$ the unique polynomial of degree $k-p+1$ in $y$ such that ${\bf P}(i)=P_{p+i}^k$ for each $i=0\dots k-p$. By linearity eq. (\ref{complete2}) gives
{\footnotesize\[
 {\bf P}(y_0)\partial_{y_0y_1}\dots\partial_{y_{k-p-1}y_{k-p}}={{\bf P}(y_1)\partial_{y_1y_2}\dots\partial_{y_{k-p-1}y_{k-p}}-{\bf P}(y_0)\partial_{y_0y_1}\dots\partial_{y_{k-p-2}y_{k-p-1}}\over y_{k-p}-y_0}.
\]
}
Specializing at $y_i=i$, one obtains
\begin{equation}\label{Pdivdif}
 {\bf P}\partial_{0\dots k-p}={{\bf P}\partial_{1\dots k-p}-{\bf P}\partial_{0\dots k-p-1}\over k-p}.
\end{equation}

By definition of ${\bf P}$ and using corollary \ref{DiffStirl}, one has
\[
{\cal B}^{-1}_{k-p}[P_p^k,\dots,P_k^k]={\cal B}^{-1}[{\bf P}(0),\dots,{\bf P}(k-p)]=(k-p)!{\bf P}\partial_{0\dots k-p}.
\]
Hence, eq. (\ref{Pdivdif}) yields
\[
{\cal B}^{-1}_{k-p}[P_p^k,\dots,P_k^k]=(k-p-1)!({\bf P}\partial_{1\dots k-p}-{\bf P}\partial_{0\dots k-p-1}).
\]
And then,
\begin{equation}\label{BtoB}
{\cal B}^{-1}_{k-p}[P_p^k, \dots, P_k^k]={\cal B}^{-1}_{k-p-1}[P^k_{p+1}, \dots, P_k^k]-{\cal B}^{-1}_{k-p-1}[P^k_p,\dots,P^k_{k-1}].
\end{equation}
Remarking that if $i<k$ one has
\[
P_i^k(x;a,b)=(x+a)P_i^{k-1}(x;a+1,b)
\]
and if $i>0$
\[
P_{i+1}^{k}(x;a,b)=(x+b)P_{i}^{k-1}(x;a,b-1),
\]
eq. (\ref{BtoB}) becomes
{\footnotesize \[\begin{array}{rcl}
{\cal B}^{-1}_{k-p}[P_p^k,\dots,P_k^k]&=&(x+b){\cal B}^{-1}_{k-p-1}[P_p^{k-1}(x;a,b-1),\dots,P_{k-1}^{k-1}(x;a,b-1)]\\&&-
(x+a){\cal B}^{-1}_{k-p-1}[P_p^{k-1}(x;a+1,b),\dots,P_{k-1}^{k-1}(x;a+1,b)].
\end{array}\]}
One recovers the result by applying the induction hypothesis.$\Box$\\
Note that, setting $p=0$ in the previous proposition, one obtains
\begin{equation}\label{LeadCoeffk}
{\cal B}_k^{-1}[P_0^k,\ldots, P_k^k]=\prod_{i=0}^{k-1}(b-a+i).
\end{equation}
\begin{remark}\label{rdegree}
Remark \ref{Matthieu} and proposition \ref{LeadCoeff} imply that the (Newton) polynomial interpolating the point $(0,P_0^k), \dots, (k,P_k^k)$ ({\emph i.e.} the polynomial {\bf P} of degree $k$ in $y$ such that ${\bf P}(x,y)=P_i^k$) is a bi-variate polynomial (in $x$ and $y$)  whose degree is $k$.
\end{remark}

\subsection{Leading coefficients}
Let ${\bf f}(x,y)$ be a bivariate polynomial of degree $m$. In this section, we investigate the coefficient of the leading term of the binomial transform ${\cal B}_k^{-1}[({\bf f}(x,i)i^p)_i]$, that is the scalar
\[
 {\cal L}_{k,p}({\bf f})=\left.[x^{p+m-k}]{\bf f}(x,y_0)y_0^p\partial_0\dots\partial_{k-1}\right|_{y_i=i}
\]
multiplied by $k!$, where $[x^i]P(x)=\alpha_i$ if $P(x)=\sum_i\alpha_ix_i$.
The following result explains how to manage the shift induced by the multiplication by $i^p$:
\begin{proposition}\label{y->y+p}
One has $${\cal L}_{k,p}({\bf f})=\left\{
 \begin{array}{ll}
{\cal L}_{k-p,0}({\bf f}_p)&\mbox{ if }p\leq k\\
0&\mbox{ otherwise,}
\end{array}
\right.$$
where ${\bf f}_p(x,y)={\bf f}(x,y+p)$.
\end{proposition}
{\bf Proof} Proposition
  \ref {DiffShift} yields
$$
{\bf f}(x,y)y^{p}\partial_{0\cdots k+1}=
{\bf f}(x,y+1)(y+1)^{p-1}\partial_{0\cdots k}.$$ 
Note that, since ${\bf f}(x,y)(y+1)^p$ is a bivariate polynomial of degree $p+m-1$,
$$
[x^{p+m-k}]{\bf f}(x,y+1)(y+1)^{p-1}=[x^{p+m-k}]f(x,y+1)(y)^{p-1} + R(y),
$$
where $R$ is a polynomial of degree at most $k-1$.
Hence, since the operator $\partial_{y_0y_1}\cdots \partial_{y_{k-1}y_k}$ lowers the degree of $k$ in the $y_i$'s, one observes
$$
[x^{p+m-k}]{\bf f}(x,y+1)(y+1)^{p-1}\partial_{0\cdots k}=
[x^{p+m-k}]
{\bf f}(x,y+1)(y)^{p-1}\partial_{0\cdots k},$$
or, equivalently,
\[{\cal L}_{k,p}({\bf f})=\left\{
 \begin{array}{ll}
{\cal L}_{k-1,p-1}({\bf f}_1)&\mbox{ if }1\leq k\\
0&\mbox{ otherwise,}
\end{array}
\right.\] 
Iterating the process, one proves the claim.
${\Box}$

If $(a_i(x))_i$ is a sequence of polynomials, one defines
\[
 \wt{\cal T}_k^{a,b}[(a_i(x))_{i}]:=(-1)^k{\cal
B}_k^{-1}\left[\left(P_i^k a_i(x)\right)_{i}\right].
\]

\begin{remark}\label{Tdegree}
From remark \ref{rdegree} if ${\bf f}(x,y)$ is a bivariate polynomial of degree $m$, its transform, $\wt{\cal T}_k^{a,b}[{\bf f}(x,0),\dots,{\bf f}(x,k)]$, is a polynomial in $x$ whose degree is also (at most) $m$. Indeed, the polynomial $\bf F$ of degree $k$ in $y$ interpolating the points $(0,f(x,0)P_0^k), \dots, (k,f(x,k)P_k^k)$ is a bivariate polynomial whose degrees in $x$ and in $\{x,y\}$ are $m+k$. Acting by $\partial_{0\dots k}$, one obtains a polynomial in $x$ which is a linear combination of coefficients $[y^i]{\bf F}$ (whose  degree in $x$ is $m+k-i$). We conclude by noting that, when $i<k$, $[y^i]{\bf F}$ gives no contribution since  $y_0^i\partial_{y_0y_1}\dots \partial_{y_{k-1}y_k}$ is a polynomial of degree $i-k$.
\end{remark}

In particular, when ${\bf f}(x,y)=y^p$, eq. (\ref{LeadCoeffk}) and proposition \ref{y->y+p} imply
\begin{equation}\label{LeadingT}
[x^p]\wt{\cal T}_k^{a,b}[(i^p)_{i}]=
\left\{\begin{array}{ll} (-1)^k\displaystyle{{k!\over
(k-p)!}\prod_{i=0}^{k-p-1}(b-a-p-i)}&\mbox{ if }p\leq k\\
0&\mbox{ otherwise}.\end{array}\right.
\end{equation}
The following result allows us to compute the leading term in the action of $\wt{\cal T}_k^{a,b}$ on a product of linear factors of the  form $cx-dy+e$.
\begin{corollary}\label{CorleadingT}

{\footnotesize
\[
\begin{array}{l}\displaystyle
[x^p]{\cal
T}_k^{a,b}\left[\left(\displaystyle\prod_{j=0}^{p-1}\left(c_jx-d_ji+e_j\right)\right)_i\right]
=\\\displaystyle \sum_{j=0}^{k}{k!\over
(k-j)!}\left(\displaystyle\sum_{\{0,\dots,p-1\}=\atop\{s_1,\dots,s_{j}\}\cup
\{t_1,\dots,t_{p-j}\}}d_{s_1}\dots d_{s_j}c_{t_1}\dots c_{t_{p-j}}
\right)\prod_{i=0}^{k-j-1}(a-b+j+i).\end{array}
\]
}
\end{corollary}
 {\bf Proof}
 First note that the variables $e_i$ give no contribution:
\[
[x^p]{\cal
T}_k\left[\left(\displaystyle\prod_{j=0}^{p-1}\left(c_jx-d_ji+e_j\right)\right)_i\right]
=[x^p]{\cal
T}_k\left[\left(\displaystyle\prod_{j=0}^{p-1}\left(c_jx-d_ji\right)\right)_i\right].
\]

Indeed, the coefficient of $e_{j_1}\dots e_{j_s}$ in
$\prod_{j=0}^{p-1}\left(c_jx-d_jy+e_j\right)$ is a polynomial
$f(x,y)$ whose degree is $p-s$. From remark \ref{Matthieu}, the
degree of ${\cal T}_k\left[\left(f(x,i)\right)_i\right]$ being at
most $p-s$, it follows immediately that $[x^p]{\cal
T}_k\left[\left(f(x,i)\right)_i\right]\neq0$ only if $s=0$.

Now, to obtain our result, it suffices to expand
$\prod_{j=0}^{p-1}\left(c_jx-d_jy\right)$ as a polynomial in $x$
and $y$:
\[\begin{array}{l}\displaystyle
\ [x^p]{\cal
T}_k\left[\left(\displaystyle\prod_{j=0}^{p-1}\left(c_jx-d_ji+e_j\right)\right)_i\right]
=\\
\displaystyle [x^p]\sum_{j=0}^p(-1)^{p-j}
\left(\displaystyle\sum_{\{0,\dots,p-1\}=\atop\{s_1,\dots,s_{j}\}\cup
\{t_1,\dots,t_{p-j}\}}c_{s_1}\dots c_{s_j}d_{t_1}\dots d_{t_{p-j}}
\right){\cal T}_k\left[\left(x^ji^{p-j}\right)_{i}\right]\\
\displaystyle=\sum_{j=0}^p(-1)^{p-j}\left(\displaystyle\sum_{\{0,\dots,p-1\}=\atop\{s_1,\dots,s_{j}\}\cup
\{t_1,\dots,t_{p-j}\}}c_{s_1}\dots c_{s_j}d_{t_1}\dots d_{t_{p-j}}
\right)[x^{p-j}]{\cal
T}_k\left[\left(i^{p-j}\right)_{i}\right].\end{array}
\]
Applying  eq. (\ref{LeadingT}), one recovers our
result.$\Box$

In the next section, we use the results of this section in the aim to investigate the asymptotic behavior of the integrals $I_k\over N$. In particular, we  show that its convergence (section \ref{Asymp}) is a direct consequence of remark \ref{Tdegree}  and  we compute explicitly  the limit (section \ref{limit}) by means of corollary \ref{CorleadingT}.
\section{Asymptotic behavior of $I_k\over N$\label{Asymp}}
In this section, we use the tools described in the previous section to prove the convergence of
the integral $I_k\over N$ and compute the limit.
\subsection{Convergence\label{convergence}} Suppose now that $a=a(N)$ and
$b=b(N)$ are linear function of $N$.  One has:
\begin{theorem}\label{CVI_k}
\[
 \left|\lim_{N\rightarrow\infty}{I_k\over N}\right|<+\infty
\]
\end{theorem}
{\bf Proof} We start from Corollary \ref{I_k} and write
\[
{I_k\over N}=\frac1{k!}{{\cal N}_k(N)\over
N\prod_{j=-k+1}^{k-1}(a(N)+b(N)+2N+j-2)}
\]
where {\footnotesize
\begin{equation}\label{refN}
\begin{array}{l}
{\cal N}_k(N):=\displaystyle\sum_{i=0}^{k-1}(-1)^i\left(k-1\atop
i\right)\prod_{j=-k+1}^{-i-1}(2N+a(N)+b(N)+j-2)\times\\\times
\displaystyle\prod_{j=-i}^{k-1-i}(N+j)(N+a(N)+j-1)\prod_{j=k-i}^{k-1}(2N+a(N)+b(N)+j-2).\end{array}
\end{equation}}
We need the following lemma:
\begin{lemma}\label{degN}
The degree in $N$ of the polynomial ${\cal N}_k(N)$ is $2k$.
\end{lemma}
{\bf Proof} For convenience, set $a=a_1N+a_0$ and $b=b_1N+b_0$,
and write
{\footnotesize
\[
\begin{array}{l}\displaystyle
\prod_{j=-k+1}^{-i-1}(2x+a(x)+b(x)+j-2)\prod_{j=k-i}^{k-1}(2x+a(x)+b(x)+j-2),\,
=\\\displaystyle\prod_{j=0}^{k-i-2}((2+a_1+b_1)x+j+a_0+b_0-1-k)\prod_{j=0}^{i-1}((2+a_1+b_1)x
+a_0+b_0-j+k-3).\end{array}\]} With the notation of the previous
section, one recognizes
{\footnotesize
\[\begin{array}{l}\displaystyle\prod_{j=0}^{k-i-2}((2+a_1+b_1)x+j+a_0+b_0-1-k)\prod_{j=0}^{i-1}((2+a_1+b_1)x
+a_0+b_0-j+k-3)=\\P_i^{k-1}((2+a_1+b_1)x;a_0+b_0-k-1,a_0+b_0+k-3).\end{array}
\]}
 If one sets $${\bf
Q}_k(x,y):=\prod_{j=0}^{k-1}\left({x\over 2+a_1+b_1}+j-y\right)\left({1+a_1\over
2+a_1+b_1}x+a_0+j-1-y\right),$$
the following holds \begin{equation}\label{N2T} {\cal N}_k(N)={\wt {\cal
T}}^{a_0+b_0-1-k,a_0+b_0+k-3}_{k-1}\left[({\bf
Q}_k(x,i))_{i\in\N}\right]|_{x=(a_1+b_1+2)N}.
\end{equation}
Hence, from remark \ref{Tdegree}, the degree of ${\cal N}_k(N)$
equals the degree of ${\bf Q}_k(x,y)$ that is $2k$.
$\Box$.\\ \\
 The degree in $N$ of
the denominator $$N\prod_{j=-k+1}^{k-1}(a(N)+b(N)+2N+j-2)$$ of
$I_k\over N$ is $2k$. From lemma \ref{degN}, $I_k\over N$ is a
rational fraction in $N$ whose  numerator and  denominator
have the same degree $2k$. Hence $I_k\over N$ converges.$\Box$
\subsection{Computation of the limit \label{limit}}
Let ${\bf
F}_{\alpha_1,\beta_1;\alpha_2,\beta_2}^p(x,y)=\prod_{j=0}^{p-1}(\alpha_1x+j-y+\beta_1)(\alpha_2x+j-y+\beta_2)$.
And set, for convenience, ${\mathbb
F}_{\alpha_1,\beta_1;\alpha_2,\beta_2}^p(x)=\left({\bf
F}_{\alpha_1,\beta_1;\alpha_2,\beta_2}^p(x,i)\right)_{i\in\N}$
One has:

\begin{proposition}\label{coeffbeta}
The coefficient of $x^{2p}$ in $\wt{\cal T}_k^{a,b}[{\mathbb
F}_{\alpha_1,\beta_1;\alpha_2,\beta_2}^p(x)]$ does not depend on
$\beta_1$ and $\beta_2$. More precisely, one has: {\footnotesize
\[
[x^{2p}]\wt{\cal T}_k^{a,b}[{\mathbb
F}_{\alpha_1,\beta_1;\alpha_2,\beta_2}^{p}(x)]=
\sum_{j=0}^k{k!\over(k-j)!}\sum_{i=0}^{p}\left(p\atop
i\right)\left(p\atop
2p-j-i\right)\alpha_1^i\alpha_2^{2p-i-j}\prod_{i=0}^{k-j-1}(a-b+j+i).
\]}
\end{proposition}
{\bf Proof} This equality is obtained from corollary
\ref{CorleadingT} setting $c_j=\alpha_1$, $c_{j+p}=\alpha_2$,
$d_j=d_{j+p}=1$, $e_j=j+\beta_1$ and $e_{j+p}=\beta_2$ for each
$j=0,1, \dots, p-1$.
 $\Box$
\\ Using this result, one finds :
\begin{theorem}
Setting $a=a_1N+a_0$ and $b=b_1N+b_0$, one has
{\footnotesize
\begin{equation}\label{limit2}
\displaystyle \lim_{N\rightarrow\infty}{I_k\over N}=
\frac{1+a_1}{k(2+a_1+b_1)^k}\sum_{j=0}^{k-1}(-1)^j\left(1+a_1\over
2+a_1+b_1\right)^j\left(j+k-1\atop
j\right)\sum_{i=0}^{k-1-j}(1+a_1)^i\left(k\atop
i+j+1\right)\left(k\atop i\right).
\end{equation}}
\end{theorem}
{\bf Proof} We have seen (see eq (\ref{N2T})) that
\[\begin{array}{l}\displaystyle
\lim_{N\rightarrow\infty}{I_k\over
N}=\frac1{k!(a_1+b_1+2)^{2k-1}}\times\\\times\displaystyle[N^{2k}]
{\wt {\cal T}}^{a_0+b_0-1-k,a_0+b_0+k-3}_{k-1}[{\mathbb
F}^k_{{1\over 2+a_1+b_1},0;{1+a_1\over
2+a_1+b_1},a_0-1}(x)]|_{x=(2+a_1+b_1)N}.
\end{array}
\]
 By proposition \ref{coeffbeta}, one obtains
\[\begin{array}{l}\displaystyle
\lim_{N\rightarrow\infty}{I_k\over N}={2+a_1+b_1\over
k}\sum_{i=0}^{k-1}(1+a_1)^{-i}{\displaystyle\prod_{j=0}^{k-i-2}(2(1-k)+j+i)}{1\over
(k-i-1)!}\times\\\displaystyle\times\sum_{j=0}^k\left(k\atop
j\right)\left(k\atop 2k-i-j\right)\left(1+a_1\over
2+a_1+b_1\right)^{2k-j}.\end{array}
\]
One recognizes
\[
{\displaystyle\prod_{j=0}^{k-i-2}(2(1-k)+j+i)}{1\over
(k-i-1)!}=(-1)^{i+k-1}\left(2(k-1)-i\atop k-1\right).
\]
And, slightly  rearranging the terms of the sum, we show the
 theorem.$\Box$\\ \\
Let us  rewrite (\ref{limit2}) as
 {\footnotesize
\begin{equation}\label{limit1.1}
\begin{array}{rcl}
\displaystyle \lim_{N\rightarrow\infty}{I_k\over N}&=&
\frac{1+a_1}{(2+a_1+b_1)^k}\left((1+a_1)^{k-1}+\right.\\
&&\left.\displaystyle\sum_{i=0}^{k-2}{(1+a_1)^i\over
k-i-1}\left(k\atop
i\right)\sum_{j=0}^{k-1-i}(-1)^j\left(1+a_1\over
2+a_1+b_1\right)^j\left(j+k-1\atop i+j+1\right)\left(k-i-1\atop
j\right)\right)\end{array}.
\end{equation}}
For convenience, we set $a_1=\ell_1-1$ and
$b_1=\frac1{\ell_2}-\ell_1-1$. With this notation, equation
(\ref{limit1.1}) reads {\footnotesize
\[
\displaystyle \lim_{N\rightarrow\infty}{I_k\over N}=
{\ell_1}{\ell_2^k}\left(\ell_1^{k-1}+\displaystyle\sum_{i=0}^{k-2}{\ell_1^i\over
k-i-1}\left(k\atop
i\right)\sum_{j=0}^{k-1-i}(-1)^j(\ell_1\ell_2)^j\left(j+k-1\atop
i+j+1\right)\left(k-i-1\atop j\right)\right).
\]
} Rearranging the sum, one obtains {\footnotesize
\begin{equation}\label{limit1}
 \displaystyle \lim_{N\rightarrow\infty}{I_k\over N}=
 \ell_1\ell_2^k\left(1+\sum_{j=1}^{k-1}{\ell_1^j\over j}\left(k\atop j+1\right)
 \left(\sum_{i=1}^{j}(-1)^j\left(j\atop i\right)\left(i+k-1\atop j-1\right){\ell^i_2}\right)
 \right),
 \end{equation}
} or equivalently, in terms of binomial transform one has
\begin{equation}
 \displaystyle \lim_{N\rightarrow\infty}{I_k\over N}=
 \ell_1\ell_2^k\left(1+\sum_{j=1}^{k-1}{\ell_1^j\over j}\left(k\atop j+1\right)
 {\cal B}_j^{-1}\left[\left(\left(i+k-1\atop
 j-1\right){\ell^i_2}\right)_i\right]
 \right).
 \end{equation}
\section{Some special cases related to combinatorics\label{Combi}}
\subsection{Simplest cases}
\begin{corollary}
\begin{enumerate}
\item If $a_1=-1$ and $b_1\neq -1$ then
$\lim_{N\rightarrow\infty}{I_k\over N}=0$. \item If $a_1\neq 1$
and $b_1=-1$ then $\lim_{N\rightarrow\infty}{I_k\over N}=1$.
\end{enumerate}
\end{corollary}
{\bf Proof} The assertion (1) is straightforward from
(\ref{limit2}).\\To prove the second assertion, we need the following
lemma
  \begin{lemma}\label{bibident1}
 Let $a,\, b$ and $c$ be three integers and denote
 \[
\left\{c\atop b\right\}_a:=\sum_{j=0}^a(-1)^j\left(a\atop
j\right)\left(c+j\atop b+j\right).
 \]
 With this notation, one has
 \[
 \left\{c\atop b\right\}_a=(-1)^a\left(c\atop a+b\right).
 \]
  \end{lemma}
  {\bf Proof} By induction, remarking that
  \[
\left\{c\atop b\right\}_a=\left\{c-1\atop
b\right\}_a+\left\{c-1\atop b-1\right\}_a.
  \]
  $\Box$.\\
 Under the specialization $b_1=-1$ (or equivalently $\ell_2=\ell_1$) and using the notation of lemma
 \ref{bibident1}, formula (\ref{limit1}) reads
 \[
 \lim_{N\rightarrow\infty}{I_k\over N}={\ell_1^{k-1}}\left(
 \ell_1^{1-k}+\sum_{i=0}^{k-2}{\ell_1^{-i}\over k-i-1}\left(k\atop
 i\right)\left\{k-1\atop i+1\right\}_{k-i-1}\right).
 \]
 But lemma \ref{bibident1} yields $\left\{k-1\atop
 i+1\right\}_{k-i-1}=\left(k-1\atop k\right)=0$. This implies our
 result.
$\Box$.
\subsection{Central binomial coefficients and the specialization $a_1=b_1=0$}
 Under this specialization,
(\ref{limit2}) reads
\[
 \lim_{N\rightarrow\infty}{I_k\over
 N}=\frac1{2^kk}\sum_{j=0}^{k-1}\left(-1\over
 2\right)^i\left(j+k-1\atop
 j\right)\sum_{i=0}^{k-1-j}\left(k\atop i+j+1\right)\left(k\atop
 i\right).
\]
But, one has
\begin{equation} \label{eq1a=0b=0}
\sum_{i=0}^{k-1-j}\left(k\atop i+j+1\right)\left(k\atop
 i\right)=\left(2k\atop k+j+1\right).
\end{equation}
Indeed, formula (\ref{eq1a=0b=0}) follows from a well known
equality
\[
 \sum_{j=0}^{\infty}\left(a\atop j\right)\left(b\atop
 c+j\right)=\left(a+b\atop a+c\right),
\]
which can be proved by a straightforward induction on $b$. Hence,
from (\ref{eq1a=0b=0}), one obtains
\[
 \lim_{N\rightarrow\infty}{I_k\over
 N}=\frac1{2^kk}\sum_{j=0}^{k-1}\left(-1\over
 2\right)^i\left(j+k-1\atop
 j\right)\left(2k\atop k+j+1\right).
\]
We need the following lemma :
\begin{lemma}\label{bibident2}
One has
\[
\left\langle n\atop
m\right\rangle:=\sum_{j=0}^{n-m}(-2)^{-j}\left(n-m+j\atop
j\right)\left(2n\atop n+m+j\right)=2^{m-n}\left(n\atop
m\right){\left(2n\atop n\right)\over\left(2m\atop m\right)}.
\]
\end{lemma}
{\bf Proof}
 Set $F_{n,j}:=(-2)^{-j}\left(n-m+j\atop j\right)\left(2n\atop
 n+m+j\right)$. We compute the associated Gosper sequence:
 \[
 G_{n,j}:=-2{j(2n+1)\over m-n+j-1}F_{n,j},
 \]
 and we check that the two sequences verify the equality
 \[
 (2n+1)F_{n,j}+(m-n-1)F_{n+1,j}=G_{n,j+1}-G_{n,j}.
 \]
 It follows that
 \[
 (2n+1)\left\langle n\atop
m\right\rangle=(n-m+1)\left\langle n+1\atop m\right\rangle.
 \]
Hence, the result is obtained by induction.$\Box$\\ \\
From \ref{bibident1}, one obtains
\begin{corollary}
\[
 \lim_{N\rightarrow\infty}{I_k\over
 N}=\frac1{2^kk}\left\langle k\atop
1\right\rangle=\frac{1}{2^{2k}}\left(2k\atop k\right).
 \]
\end{corollary}

This result is well-known in the quantum scattering setting in symmetric cavities \cite{Nov1,Nov2,Vivo}.

\subsection{Catalan triangle and the specialization $a_1=0$}
 Let us set for
convenience $b_1=\ell-1$. Under this specialization eq
(\ref{limit2}) reads
\[
\lim_{N\rightarrow\infty}{I_k\over N}={1\over
k(1+\ell)^k}\sum_{j=0}^{k-1}(-1)^j\left(1\over
1+\ell\right)^{j}\left(j+k-1\atop
j\right)\sum_{i=0}^{k-1-j}\left(k\atop i+j+1\right)\left(k\atop
i\right).
\]
Using formula (\ref{eq1a=0b=0}), we obtain
\[
\lim_{N\rightarrow\infty}{I_k\over N}={1\over
k(1+\ell)^k}\sum_{j=0}^{k-1}(-1)^j\left(1\over
1+\ell\right)^{j}\left(j+k-1\atop j\right)\left(2k\atop
k+j+1\right),
\]
or equivalently {\footnotesize
$$\begin{array}{rcl}
\displaystyle \lim_{N\rightarrow\infty}{I_k\over
N}&=&\displaystyle{1\over
k(1+\ell)^{2k-1}}\sum_{j=0}^{k-1}(-1)^j\left(
1+\ell\right)^{k-j-1}\left(j+k-1\atop j\right)\left(2k\atop
k+j+1\right)\\
&=& \displaystyle {1\over
k(1+\ell)^{2k-1}}\sum_{i=0}^{k-1}\left(\sum_{j=0}^{k-1-i}(-1)^j\left(k-1-j\atop
i\right)\left(k+j-1\atop j\right)\left(2k\atop
j+k+1\right)\right)\ell^i.\end{array}
$$
} By rearranging the factor of the products appearing in the
coefficient of each $\ell^i$, one restates this expression in
terms of inverse binomial transform:
\begin{equation}\label{INbinom}
\lim_{N\rightarrow\infty}{I_k\over N}= {(2k)!\over
k!(1+\ell)^{2k-1}}\sum_{i=0}^{k-1}{(-1)^{k-1-i}\over
i!(k-i-1)}{\cal B}^{-1}_{k-i-1}\left[\left(1\over
(j+k)(j+1+k)\right)_j\right]\ell^i
\end{equation}
We need the following lemma:
\begin{lemma}\label{bintranslem3}
 \[
 {\cal B}_m^{-1}\left[\left(1\over (p+i)(p+i+1)\right)_i\right]={(-1)^m(m+1)!\over
 \prod_{i=0}^{m+1}(p+i)}.
 \]
\end{lemma}
{\bf Proof}
 First remark that
 \[
{ 1\over (p+i)(p+i+1)}={P_{m-i}^m(p;0,m+1)\over
 \prod_{i=0}^{m+1}(p+i)}.
 \]
 Hence,
\[
 {\cal B}_m^{-1}\left[\left(1\over (p+i)(p+i+1)\right)_i\right]={{\cal B}_m^{-1}\left[P^m_0(p;0,m+1)\dots 
P^m_m(p;0,m+1)\right]\over
 \prod_{i=0}^{m+1}(p+i)}.
 \]
 We conclude by using eq. (\ref{LeadCoeffk}).
$\Box$\\ \\
%
%
%
Now using lemma \ref{bintranslem3}
 in equality (\ref{INbinom}), one finds
\begin{proposition}
\[
\lim_{N\rightarrow\infty}{I_k\over
N}={\displaystyle\sum_{i=0}^{k-1}{k-i\over k}\left(2k\atop
i\right)\ell^i\over (1+\ell)^{2k-1}}
\]
\end{proposition}

\noindent The triangle $\mathbb{a}:=\left({k-i\over
k}\left(2k\atop i\right)\right)_{k,i\in\N}$ is sometimes called
Catalan triangle (see {\it e.g.} sequences {\tt A008315}, {\tt
A050166} and {\tt
A039598} in \cite{Sloane}).\\
 Note that these numbers are related
to many combinatorial objects. For example, they appear in the
expansion of odd power of $x$ in terms of orthogonal Chebyshev
polynomials $U_k(x)$ of the second kind (see {\it e.g.}
\cite{Handbook} p.796), since
\[
 x^{2k-1}=\frac1{2^{2k-1}}\sum_{i=0}^{k-1}{\mathbb a}_{k,i}U_{2(k-i)-1}(x).
\]
Another example is given by  R.K. Guy in \cite{Guy}: he showed
that the number of walks in a lattice with $k$ steps (each in
direction N, S, E or W) starting at $(0,0)$ and at a distance $i$
from the $x$-axis equals ${\mathbb a}_{k+1-i,k+1}$.

\subsection{Symmetric Dyck paths counted by number of peaks and the specialization
$b_1=0$}



For convenience, let us set $a_1=\ell-1$. This case has been
already computed by M. Novaes in \cite{Nov1,Nov2}. With our
notation, he proved the following formula
\begin{equation}\label{novaes}
\lim_{N\rightarrow\infty}{I_k\over
N}=(\ell+1)\sum_{i=1}^{k}{(-1)^{i-1}\over i}\left(k-1\atop
i-1\right)\left(2(i-1)\atop i-1\right)\left(\ell\over
(1+\ell)^2\right)^i. \end{equation}

Our goal is to identify the coefficient $\alpha _{i,k}$ such that
\[
\lim_{N\rightarrow\infty}{I_k\over N}=
{\sum_{i}\alpha_{i,k}\ell^i\over (1+\ell)^{2k-1}}.\]

Let us sketch the proof of the following result, announced by two of us in \cite{LV}.
\begin{proposition}
\[
 \lim_{N\rightarrow 0}{I_k\over
 N}={\ell\over(1+\ell)^{2k-1}}\sum_{i=0}^{2(k-1)}\left(k-1\atop \left\lceil i\over2\right\rceil\right)
 \left(k-1\atop \left\lfloor i\over2\right\rfloor\right)\ell^i
\]
\end{proposition}
{\bf Proof}
 First rewrite eq (\ref{novaes}) as
 \[
\lim_{N\rightarrow\infty}{I_k\over
N}=\frac1{2k-1}\sum_{i=1}^{k}{(-1)^{i-1}\over i}\left(k-1\atop
i-1\right)\left(2(i-1)\atop i-1\right)\ell^i\left(
1+\ell\right)^{2(k-i)}.
 \]
Expanding $\left( 1+\ell\right)^{2(k-i)}$ and rearranging the sum
one obtains
\[
\lim_{N\rightarrow\infty}{I_k\over N}={\ell\over
(1+\ell)^{2k-1}}\sum_{i=0}^{2(k-1)}\left(\sum_{j=0}^{k-1}(-1)^j\left(2(k-j-1)\atop
i-j\right){\left(2j\atop j\right)\over j+1}\right)\ell^i
\]
Hence, we need to prove:
\[
{\gimel}_{k,i}:=\sum_{j=0}^{k}(-1)^j\left(2(k-j)\atop
i-j\right){\left(2j\atop j\right)\over j+1}=\left(k\atop
\left\lceil i\over2\right\rceil\right)
 \left(k\atop \left\lfloor i\over2\right\rfloor\right).
\]
Using Gosper algorithm and the Zeilberger method (see {\it e.g.}
\cite{A=B}), we find that the sequence ${\mathbb \gimel}_{k,i}$ is
completely determined by the following three terms relation:
\begin{equation}\label{threeterms}-\left (2\,k+1-i\right )\left (2\,k-i\right
)\gimel_{k,i}+\left (2+2\,i-2\,k \right )\gimel_{k,i+1}+\left
(i+3\right )\left (i+2\right) \gimel_{k,i+2}=0,\end{equation} and
the initial conditions
\begin{equation}\label{initial}
{\gimel}_{k,0}=1,\,{\gimel}_{k,1}=k.
\end{equation}
A straightforward calculation shows that the numbers $\left(k\atop
\left\lceil i\over2\right\rceil\right)
 \left(k\atop \left\lfloor i\over2\right\rfloor\right)$ verify
 also eq (\ref{threeterms}) and (\ref{initial}). This concludes
 the proof.
$\Box$
\\ \\
Remark that the numbers $\gimel_{k,i}$ have an interesting
combinatorial interpretation since it is the number of Dyck paths
of odd semi-length $2k-1$ with $i$ peaks (see {\it e.g.}
\cite{Barry} and sequence A088855 in \cite{Sloane}).

\section{Conclusion \label{conclusion}}
 This work is the first step towards a combinatorial
interpretation for the asymptotic behavior of the integrals
\begin{equation}\label{goal}\langle\lambda\rangle_N^\sharp:=\langle x_1^{\lambda_1}\dots x_N^{\lambda_N}\rangle^\sharp
:={\langle x_1^{\lambda_1}\dots
x_N^{\lambda_N}\rangle_{a,b}^c\over \langle 1\rangle_{a,b}^c}
\end{equation}for any partition $\lambda$ where
\[
\langle f(x_1,\dots,x_N)\rangle_{a,b}^c:={\displaystyle
\int_{[0,1]^N}
f(x_1,\dots,x_N)\prod_{i<j}|x_i-x_j|^{2c}\prod_ix_i^{a-1}(1-x_1)^{b-1}dx_i
}.
\]
In particular, our goal is to find an algebraic proof of the
following conjecture suggested by numerical evidences\footnote{ At the time of writing, we are aware of a proof, based on other principles and deferred to a 
forthcoming paper, which contains eq. 18 as a special case and allows us to relate it to the combinatorics of symmetric Dyck paths.}.
\begin{conjecture}\label{conj1}
One has:
\begin{equation}\label{pgoal}\lim_{N\rightarrow\infty}\langle
\frac1{N^{\ell(\lambda)}}p_\lambda(x_1,\dots,x_N)\rangle^\sharp=\prod_{i=1}^{\ell(\lambda)}
\lim_{N\rightarrow\infty}\langle
\frac1Np_{\lambda_i}(x_1,\dots,x_N)\rangle^\sharp,\end{equation}
where $p_\lambda$ stands for the product $p_{\lambda_1}\dots
p_{\lambda_{\ell(\lambda)}}$.
\end{conjecture}
The limitations of the method described in this article is that we
need an explicit expression for the integrals. This quickly
becomes tedious in the general case.
For testing equality (\ref{pgoal}), we first obtain an expression
of a product of power sums in terms of Jack polynomials, we
implement an identity similar to (\ref{SchurEq}) for the integral
associated to each of Jack polynomials and finally we take the
limit after simplifying the sum.\\
The link between conjecture \ref{conj1} and the integral
(\ref{goal}) is the following. First, two of the authors remarked
in \cite{LV} that integral (\ref{goal}) can be restated in terms
of monomial symmetric functions
\[
\langle\lambda\rangle^\sharp_N= {\langle
m_\lambda\rangle^\sharp\over P_\lambda(N)},
\]
where $P_\lambda(N)$ is an explicit polynomial whose degree is
exactly $\ell(\lambda)$. Remarking that
\[
 m_\lambda=\alpha_\lambda
 p_\lambda+\sum_{\ell(\mu)<\ell(\lambda)}\alpha_\mu p_\mu,
\]
and assuming conjecture \ref{conj1}, we see that
$\lim_{N\rightarrow\infty}\langle\lambda\rangle^\sharp_N$ equals
(up to an explicit multiplicative coefficient)
$\lim_{N\rightarrow\infty}\langle
\frac1{N^{\ell(\lambda)}}p_\lambda(x_1,\dots,x_N)\rangle^\sharp$.\\ \\
\begin{acknowledgments}The authors are grateful to M. Novaes for fruitful discussions on the factorization conjecture.
This paper is partially supported by the ANR project PhysComb, ANR-08-BLAN-
0243-04.
\end{acknowledgments}

\appendix
\section{Calculation of $I_k$ (corollary \ref{I_k}) via spectral density\label{spectral}}

We sketch here the derivation of an alternative formula for $I_k$
(Corollary \ref{I_k}) via spectral density of the Jacobi ensemble.

The spectral density of the Jacobi ensemble is the marginal of the
joint density function (integrand of the Selberg integral),
defined as:
\begin{equation}
\rho_N(x_1;A,B)=N\int_0^1 dx_2\cdots dx_N
\prod_{j<k}|x_j-x_k|^2\prod_{i=1}^N x_i^{A}(1-x_i)^{B}
\end{equation}
and it is normalized to $N$, $\int_0^1 dx_1 \rho_N(x_1;a,b)=N$.

Such density is known analytically for all $N$ in terms of Jacobi
polynomials $P_j^{(A,B)}(z)$ as:
\begin{equation}
\rho_N(x;A,B)=x^A
(1-x)^B\sum_{j=0}^{N-1}c_j(A,B)[P_j^{(A,B)}(1-2x)]^2
\end{equation}
where:
\begin{equation}
c_j(A,B)=\frac{(2j+A+B+1)\Gamma(j+1)\Gamma(j+A+B+1)}{\Gamma(j+A+1)\Gamma(j+B+1)}
\end{equation}

The quantity $I_k$ in Corollary \ref{I_k} is a linear statistics
on the eigenvalues of the Jacobi ensemble. As such, it can be
computed as a 1-fold integral over the density:
\begin{equation}\label{IK}
I_k=\int_0^1 dx\ x^k\ \rho_N(x;A,B)
\end{equation}
where $A=a-1$, $B=b-1$.

The integral (\ref{IK}) can be evaluated in terms of nested finite
sums as:
\begin{equation}\label{nested}
I_k=\sum_{j=0}^{N-1}\frac{c_j(A,B)}{j!}\sum_{m,\ell=0}^j
d_{mj}(A,B)f_{mj}^{(k)}(A,B)g_{mj\ell}^{(k)}(A,B)
\end{equation}
where:
\[
\begin{array}{rcl}
d_{mj}(A,B) &=&\frac{(-j)_m\ (A+B+j+1)_m\ (A+m+1)_{j-m}}{m!}\\
f_{mj}^{(k)}(A,B) &=&\frac{\Gamma(A+j+1)\Gamma(A+k+m+1)\Gamma(B+1)}{j!\ \Gamma(A+1)\Gamma(A+B+k+m+2)}\\
g_{mj\ell}^{(k)}(A,B) &=&\frac{(-j)_\ell\ (j+A+B+1)_\ell\
(A+k+m+1)_\ell}{\ell!\ (A+1)_\ell\ (A+B+k+m+2)_\ell}
\end{array}
\]
where $(x)_n=\Gamma(x+n)/\Gamma(x)$ is a Pochhammer symbol.\\

The identity (\ref{nested}) is obtained straightforwardly by first
expanding one of the two Jacobi polynomials using the definition
in \cite{Weisstein2} and then computing the remaining integral
using formula 7.392.1 in \cite{GR}.
The equivalence between formula (\ref{nested}) and corollary 
 \ref{I_k} can then be proved by means of elementary but lengthy algebraic steps.

\end{document}